\documentclass[floats, twocolumn, nofootinbib]{revtex4}

\usepackage{graphicx}
\usepackage{dcolumn}
\usepackage{bm}
\usepackage{amsmath}
\usepackage{amsthm}
\usepackage{multirow}
\usepackage{enumerate}

\usepackage{amsfonts}
\usepackage{euscript}
\usepackage{ifthen}
\usepackage{psfrag}
\usepackage{slashed}
\usepackage{hyperref}
\usepackage{gensymb}
\usepackage{color}

\usepackage{verbatim}
\allowdisplaybreaks[4]
\newcommand{\GeV}{\text{GeV}}

\newcommand{\be}{\begin{equation}}
\newcommand{\ee}{\end{equation}}
\newcommand{\bea}{\begin{eqnarray}}
\newcommand{\eea}{\end{eqnarray}}

\begin{document}

\title{Broadband Dark Matter Axion Detection using a Cylindrical Capacitor}

\author{~\\Wenming Chen\\\emph{College of Physics and Optoelectronic Engineering, Department of Physics, Jinan University, Guangzhou 510632, China}\\~\\Yu Gao\footnotemark[1]\\\emph{Key Laboratory of Particle Astrophysics, Institute of High Energy Physics, Chinese Academy of Sciences, Beijing 100049, China}\\~\\Qiaoli Yang\footnotemark[1]\\\emph{College of Physics and Optoelectronic Engineering, Department of Physics, Jinan University, Guangzhou 510632, China}\\}


\affiliation{~}

\begin{abstract}
Cosmological axions/axion-like particles can compose a significant part of dark matter; however, the uncertainty of their mass is large. Here, we propose to search the axions using a cylindrical capacitor, in which the static electric field converts dark matter axions into an oscillating magnetic field. Due to the odd CPs, the axions couple to the electric field differently compared to the magnetic field. The axion couples to the electric field via a derivative that carries spatial information of incoming dark matter flux, while the coupling to the magnetic field depends on the dark matter density. This difference could be helpful in searching the axions and studies of the integrity of the theory, especially when the axions are very light, in which case the magnetic field-induced signal is DC-like. Orientation dependence could also be used to reduce the kinetic fluctuation-induced noise when multiple detectors operate simultaneously. In addition, a cylindrical setup shields the electric field to the laboratory and encompasses the axion-induced magnetic field within the capacitor. The induced oscillating magnetic field can then be picked up by a sensitive magnetometer. Adding a superconductor ring-coil system into the scheme can further boost the sensitivity and maintain the axion dark matter inherent bandwidth. This proposed setup could be capable of wide mass range searches.
\end{abstract}
\date{\today}

\maketitle

\renewcommand{\thefootnote}{\fnsymbol{footnote}}
\footnotetext[1]{Corresponding authors.}
\section{Introduction}
Cosmological observations indicate that 23\% of the energy density of the Universe is cold dark matter~\cite{Planck:2015fie}. There are many well-motivated cold dark matter candidates, such as the QCD axion and the axion-like particle (ALP) arising from the string theory. The QCD axion was originally proposed to solve the strong CP problem~\cite{Peccei:1977hh, Peccei:1977ur,Weinberg:1977ma, Wilczek:1977pj,vysotsky, Kim:1979if,Shifman:1979if,Zhitnitsky:1980tq,Dine:1981rt,Davidson:1981zd} and later found to be a promising dark matter candidate \cite{Preskill:1982cy,Abbott:1982af,Dine:1982ah,Sikivie:1982qv,Ipser:1983mw,berezhiani1985}. ALPs generally exist in string theory, where the extradimensions are compactified, thus creating many different axion species. Both QCD axions and ALPs have been extensively studied for their rich astrophysical and cosmological phenomena. Recently, lighter ALPs have received increased attention due to their roles in serving as wave-like dark matter, which can explain many puzzles of galactic halo structure formation~\cite{Hu:2000ke}.

The axion field $a$ couples to the electromagnetic fields as:
\be
{\cal L}_{a\gamma\gamma} = -g_{a\gamma}~ a \vec{E}\cdot \vec{B},
\label{agamma}
\ee
where $g_{a\gamma}$ is related to the Peccei-Quinn (PQ) symmetry breaking scale $f_a$ as $g_{a\gamma}= c_{\gamma}\alpha/(\pi f_a)$. $\alpha=1/137$ is the fine structure constant and $c_{\gamma}$ is a model-dependent coefficient of order one (e.g., $c_{\gamma}= -0.97$ in the KSVZ axion model and 0.36 in the DFSZ axion model). The axion also couples to fermions, such as electrons and protons, as:
$
{\cal L}_{af}={-(g_f/2 f_a)}\partial _{\mu}a\bar \psi_f\gamma ^{\mu}\gamma^5\psi_f~,
$
where $g_f$ is a model-dependent factor and $\psi_f$ is the respective fermion field.

The axion acquires a small mass from the nonperturbative instanton effects. For the QCD axions, the mass $m_a$ can be generally written as:
\be
m_a\approx 0.6\times 10^{-5}{\rm eV}\left({10^{12}~\GeV\over f_a}\right)~.
\ee
However, due to the uncertainties of string instantons, the ALP mass spans several orders of magnitude.

The axion mass $m_a$ and the PQ symmetry breaking scale $f_a$ are the two most important parameters of the axion searches. They are constrained by cosmological, astrophysical and laboratory considerations. For example, the QCD axion field received energy from the Standard Model sector during the QCD phase transition. If PQ symmetry breaking occurs after inflation, then $f_a\sim 10^{12}$GeV and $m_a\sim10^{-5}$eV, which is often called the classical window. If PQ symmetry breaking occurs before inflation, the initial axion field misalignment angle can be anthropic. In this scenario, the isocurvature perturbation of the cosmic microwave background gives additional constraints \cite{Hertzberg:2008wr,Gao:2019tqt} and typically leads to $f_a\gtrsim 10^{14}$GeV, $m_a\lesssim 10^{-7}$eV. Axion dark matter can also be created by the stochastic case \cite{Graham:2018jyp,Guth:2018hsa}, where $m_a$ ranges from $10^{-12}$eV to $10^{-4}$eV. For string theory-originated dark matter ALPs, the mass range is much larger \cite{Svrcek:2006yi,Visinelli:2018utg}. There are many experiments around the world looking for axions and ALPs \cite{Sikivie:1983ip,Sikivie:1985yu,DePanfilis:1987dk,Hagmann:1990tj,Ehret:2010mh,Tam:2011kw,Wouters:2013iya,Graham:2013gfa,Budker:2013hfa,Sikivie:2013laa,Stadnik:2013raa,Ayala:2014pea,Rybka:2014cya,Sikivie:2014lha,TheMADMAXWorkingGroup:2016hpc,Barbieri:2016vwg,Kahn:2016aff,Yang:2016zaz,Brubaker:2016ktl,Anastassopoulos:2017ftl,Abel:2017rtm,Graham:2017ivz,McAllister:2017lkb,Akerib:2017uem,Du:2018uak,Marsh:2018dlj,Zhong:2018rsr,Lawson:2019brd,Ouellet:2018beu,Arza:2019nta,Yang:2019xdz, ADMX:2021nhd}. Notably, the mass range $10^{-6}$eV$\sim10^{-10}$eV has been less explored.

 Many axion experiments are based on the electromagnetic coupling between axions and photons. For axions with a heavier mass, axion haloscopes such as the ADMX are the most capable ones that can reach the QCD axion parameter spaces within current technology. For lighter mass axions, it could be better to consider a quasistatic picture, in which when a strong static magnetic field is present, the axion-sourced current induces a detectable oscillating magnetic field~\cite{Sikivie:2013laa}.

For light axions, such as in the fuzzy dark matter regime, the induced magnetic field would be DC-like, which could make it harder to distinguish. Interestingly, the odd CP of the axions makes their couplings to the E field behave differently. The electric field couples to the axion field's derivative; thus, the source terms carry the directional information of the incoming dark matter flux, in addition to the dark matter local density information. Reorientation of the experimental apparatus could be used to distinguish the signals and the backgrounds. This difference is additionally helpful in studying the axion's fundamental properties, and it can serve as a test of the integrity of the beyond the standard model theory.


\begin{figure}
\includegraphics[scale=0.07]{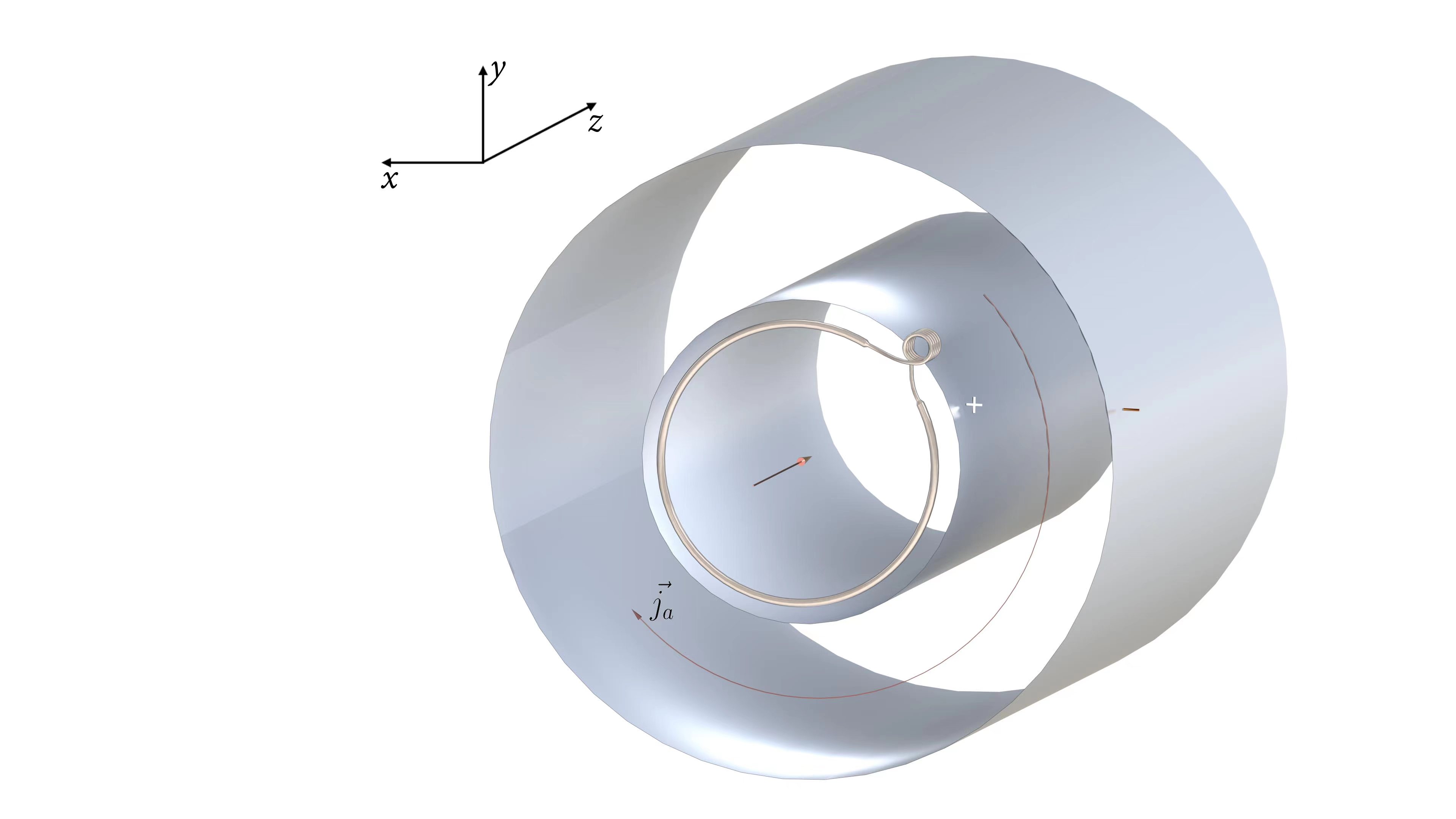}
\caption{An illustration of the proposed experiment. The electric field $\vec E$ permeates the region inside the capacitor. The dark matter axions flowing along the $\vec z$ axis give rise to an effective electric current density $\vec j_a$. The effective current density subsequently induces an oscillating magnetic field that can be picked up by a SQUID or other types of magnetometers.}
\label{fig1}
\end{figure}

\begin{figure}
\includegraphics[scale=0.06]{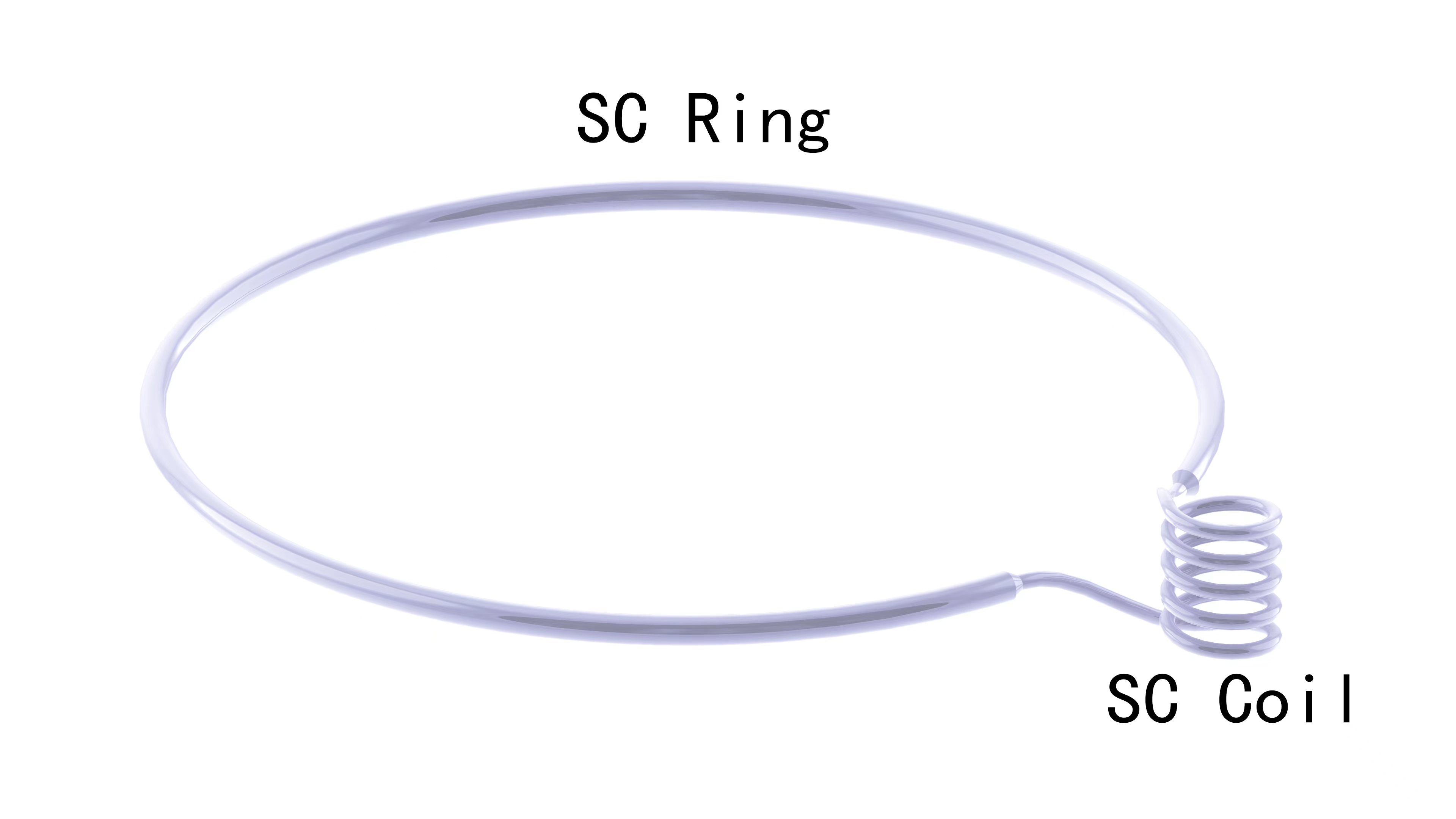}
\caption{Enhancement of the oscillating $B$-field signal can be achieved by using a superconductor ring-coil pickup. A superconductor ring couples to an $N$ turn coil, and then the magnetic field is boosted by $N*r_2 /r_1$, where $N$ is the turning of the coil. Compared to an LC circuit, a ring coil maintains a large bandwidth so does not need to be tuned to the dark matter axion frequency.}
\label{fig:pickup2}
\end{figure}

\section{Theoretical considerations and a preliminary setup}
The dark matter axions can be considered free streaming particles on the laboratory scale. In addition, the cold dark matter particles are nonrelativistic; therefore, the local axion field can be written as:
\be
a(x)\approx a_0{\rm cos}\left(-m_at-{m_a\over 2}v^2t+m_a\vec v\cdot \vec x+\phi_0\right)~,
\ee
where $a_0\approx \sqrt{2\rho_{CDM}}/m_a$ is the averaged axion field strength, $\rho_{CDM}$ is the local dark matter energy density, and $v$ is the local dark matter velocity relative to the laboratory, which we will take as $v\approx 10^{-3}$c, as the weakly interacting particles should have a speed similar to the sun in the galaxy gravitational well. $\phi_0$ is a phase factor and can be safely neglected in the following discussion. One-half of the axion de-Broglie wavelength is ($\hbar=1$)
\bea
{\lambda\over 2}={\pi\over m_av}\approx20.9{\rm ~meter}\cdot\left({10^{-5}{\rm eV}\over m_a}\right)~~.
\eea
Thus, when the axions have a mass smaller than $ 10^{-5}$ eV, the field is effectively homogeneous in the laboratory.

The axion couples to the photon via Eq. (\ref{agamma}), which gives rise to the equations of motion:
\bea
\vec\nabla\cdot \vec E&=&\rho_e+g_{a\gamma}\vec B\cdot\nabla a  \nonumber\\
\vec\nabla\times\vec B-{\partial \vec E \over \partial t}&=&g_{a\gamma}\vec E\times \vec \nabla a-g_{a\gamma}\vec B{\partial a\over  \partial t}+\vec j_e \nonumber\\
\vec\nabla\cdot\vec B&=&0\nonumber\\
\vec\nabla\times \vec E&=&-{\partial \vec B\over \partial t}~,
\eea
where $\rho_e$ and $\vec j_e$ are the ordinary electric charge density and current density, respectively. When a static electric field strength $\vec E_0$ is present, the oscillating axion field induces a small magnetic field and, up to the first order of $g_{a\gamma}$, satisfies
\bea
\vec \nabla\times\vec B=g_{a\gamma}(\vec E_0\times \vec v)\sqrt{2\rho_{CDM}}{\rm cos}\left[m_a(1+{1\over 2}v^2)t\right]= \vec j_a~,
\label{amp}
\eea
where we have assumed that the electric field permeated region is smaller than the de Broglie wavelength of the dark matter axions. Eq.(\ref{amp}) resembles Ampere's law with an axion-induced effective current density $\vec j_a$.

The dark matter axions have a small velocity distribution $\delta v$, which depends on the effective dark matter temperature $T$. The axions are generally very cold, resulting in a typical $\delta v\lesssim 10^{-7}$c \cite{Sikivie:2001fg,Armendariz-Picon:2013jej}. The velocity distribution gives rise to an inherent frequency bandwidth:
\bea
\delta f_a&=&{\delta E\over 2\pi}={m_av\over 2\pi}\delta v\sim0.5* 10^{-10}m_a\nonumber\\
&\sim& 0.75{\rm Hz}\left({m_a\over 10^{-5}{\rm eV}}\right).
\label{df}
\eea
We can define $Q_a={f_a /\delta f_a}\approx3*10^9$. $Q_a$ is the inherent quality factor of the axion dark matter.

\begin{figure}
\includegraphics[scale=0.4]{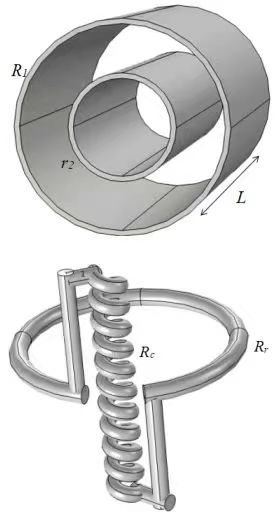}
\caption{A COMSOL modeling of the cylinder capacitor and the pickup coil.}
\label{fig:pickup}
\end{figure}

\begin{table}[t]
\begin{center}
\begin{tabular}{|c|c|p{12em}|}
		\hline\rule{0pt}{15pt}\centering
		$R_1$ &	1m & The outer radius of the shell of the cylinder  \\
		\hline\rule{0pt}{15pt}
		$r_2$ & 0.5m	& The outer radius of the inner shell of the cylinder \\
		\hline\rule{0pt}{15pt}
		$\Delta R,~\Delta r$ & 0.05m & The thickness of the shells\\
		\hline\rule{0pt}{15pt}
		$L$ & 5m & The height of the cylinder \\
		\hline\rule{0pt}{15pt}
		$R_r$ & 0.365m & The radius of the ring \\
		\hline\rule{0pt}{15pt}
		$R_c$ & 0.04m & The radius of the coil \\
		\hline\rule{0pt}{15pt}
		$N$ & 10.6 & The number of coil turns \\
		\hline\rule{0pt}{15pt}
		The Cylindrical Shell&  ~~Material& ~~Solid Silicon    \\
		\hline\rule{0pt}{15pt}
		Ring&~~Material&  ~~Gold \\
		\hline\rule{0pt}{15pt}
		Coil   & ~~Material&  ~~Gold \\
		\hline\rule{0pt}{15pt}
		Other~Areas&~~Material&  ~~Air \\
		\hline\rule{0pt}{15pt}
		Boundary~&Without *&  ~~Air \\
        \hline\rule{0pt}{15pt}
        Boundary~&With *&  ~~Copper shell \\
        \hline
\end{tabular}
\end{center}
\caption{The simulation parameters.}
\label{tab:config}
\end{table}

\begin{figure}[t]
\includegraphics[scale=0.2]{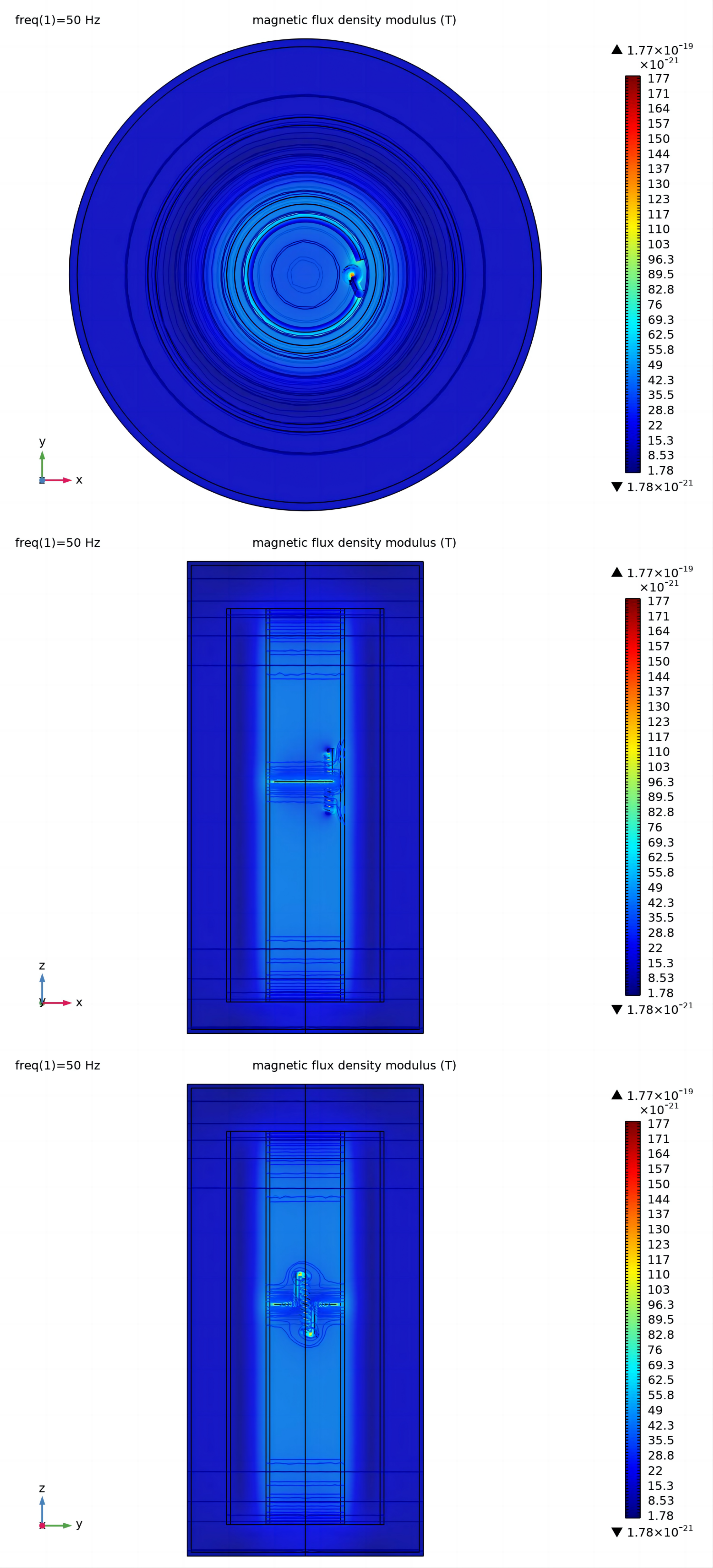}
\caption{The induced magnetic field of the scheme within a copper shell as a boundary. The axion frequency is 50 Hz.}
\label{fig:50Hz}
\end{figure}

\begin{figure}[t]
\includegraphics[scale=0.25]{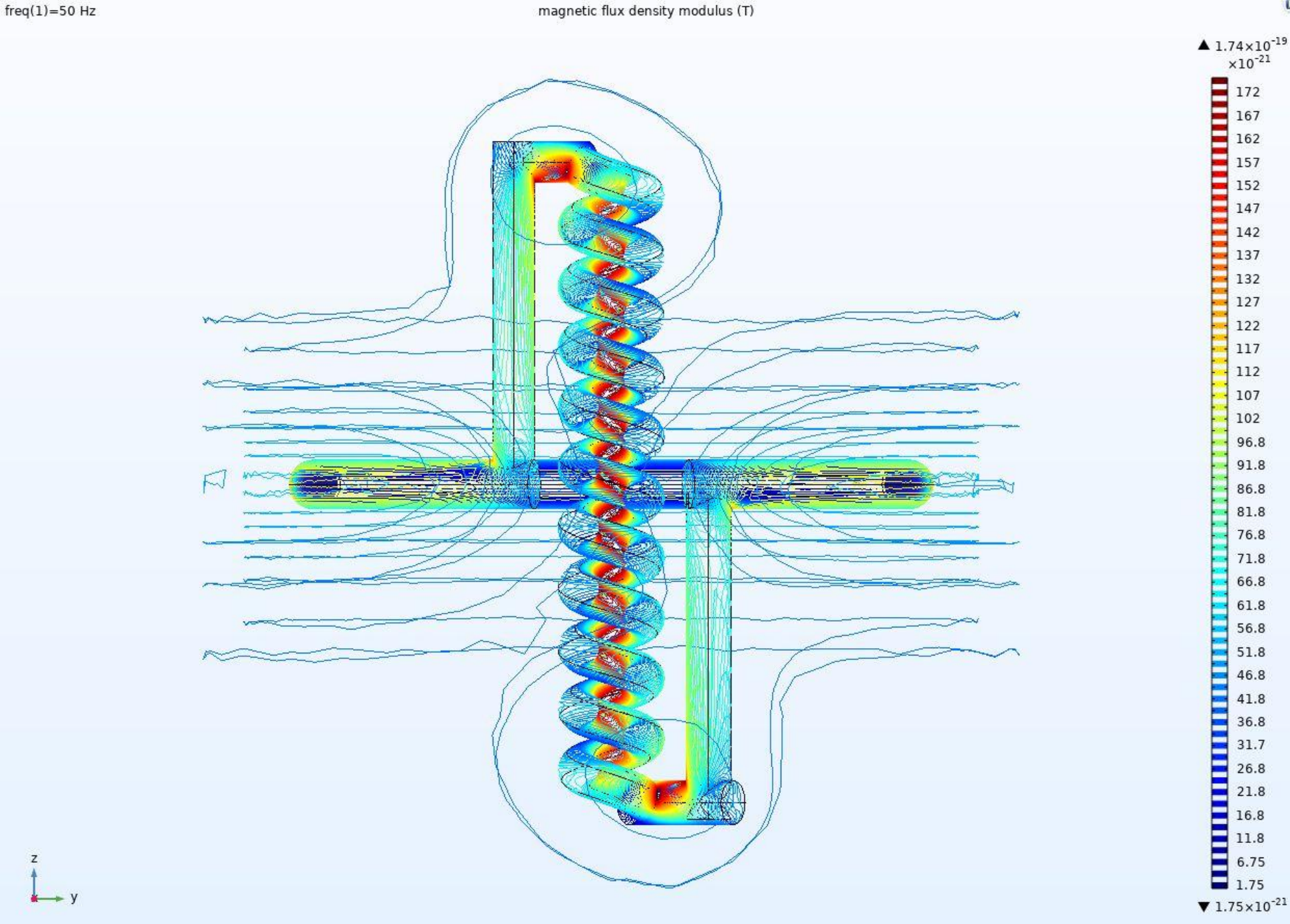}
\caption{Detailed Magnetic field distribution near the ring with the axion frequency at 50 Hz.}
\label{fig:50-2Hz}
\end{figure}

Please see Fig. (\ref{fig1}), Fig. (\ref{fig:pickup2}) and Fig. (\ref{fig:pickup}) for a preliminary experimental setup. A cylindrical capacitor can be an ideal device to create a confined region with a strong electric field $\vec E_0$. Assuming that the axile length $L$ is much larger than the radius of the cylinder, when the axion velocity $v$ is parallel to the axile, the induced effective current density $\vec j_a$ runs a closed cycle that lies in the cross-section of the cylinder. Consequently, the induced magnetic field strength $\vec{B}_a$, similar to a solenoid case, points along with the cylinder axis. Some portion of $j_a$ can be parallel to the cylinder axis when the axion velocity is partially perpendicular to the cylinder axis. The axion-induced magnetic field strength is
\bea
&B_a&=\mu_0 R j_a=g_{a\gamma}\bar E_0v\sqrt{2\rho_{CDM}}R{\rm cos}(\omega_at)\nonumber\\
&=&0.4\times 10^{-7}{\rm T}\left({g_{a\gamma}\over {\rm GeV^{-1}}}\right)\left({\bar E_0\over{\rm  Gvolt/m}}\right)\left({R\over {\rm 1m}}\right)\nonumber\\ &\times& {\rm cos}(\omega_at) \label{eq:signal_fieldstrength}
\eea
where $R$ is the distance between the two plates of the capacitor. Practically, the electric field $E_0$ is constrained by current technology, and $R$ is constrained by the axion de Broglie wavelength. Note that this result should be taken as an estimation of the order of the signals. In an actual experimental setup, one must consider the boundary condition imposed by a particular apparatus and the perturbation corrections due to the induced dynamic electric field component. Fortunately, in very light mass axion searches, the extremely small induced signal and its slow variation make materials respond weakly. We used the COMSOL package to perform 3D electromagnetic simulation for the experimental scheme with specific materials and axion frequencies, with our geometric layout shown in Fig.~\ref{fig:pickup} and configurations listed in Table~\ref{tab:config}. We find that the simulated signal $B_a$ field strength in the inner region agrees with Eq.~\ref{eq:signal_fieldstrength} toward low frequency; see Table~\ref{tab:comparison}. At higher frequencies, the screening due to the inner surface causes a larger suppression. One simulation sample is shown in Fig.~\ref{fig:50Hz},~\ref{fig:50-2Hz}. For additional simulation results at different axion frequencies, see Appendix~\ref{app:sim}.

\begin{table}
\begin{center}
\begin{tabular}{|p{5em}|c|c|c|c|}
\hline\rule{0pt}{15pt}\centering
    Freq. & Eq.~\ref{eq:signal_fieldstrength} (T) & $B_a$(T) &	$B_1$(T) & $F_r$   \\
	\hline\rule{0pt}{20pt}
	50Hz & $6.36 \times 10^{-20}$ & $2.13\times10^{-20}$ &	$1.09\times10^{-19}$ & 0.053  \\
	\hline\rule{0pt}{20pt}
	50Hz*& $6.36 \times 10^{-20}$ &$3.17\times10^{-20}$ &	$1.77\times10^{-19}$ &  0.058 \\
	\hline\rule{0pt}{20pt}
	1MHz& $6.36 \times 10^{-20}$&$1.85\times10^{-20}$ &	$1.32\times10^{-19}$ &   0.074 \\
	\hline\rule{0pt}{20pt}
	1MHz*& $6.36 \times 10^{-20}$&$2.58\times10^{-20}$ &	$2.02\times10^{-19}$ &  0.081  \\
	\hline\rule{0pt}{20pt}
	10MHz& $6.36 \times 10^{-20}$&$5.46\times10^{-21}$ &	$4.10\times10^{-20}$ &  0.077  \\
	\hline\rule{0pt}{20pt}
	10MHz* & $6.36 \times 10^{-20}$&$5.47\times10^{-21}$ &	$4.13\times10^{-20}$ &  0.078 \\	
\hline
\end{tabular}
\caption{Comparisons between Eq.~\ref{eq:signal_fieldstrength} and simulation results. $B_1$ is the amplified signal field strength inside the pickup coil. Due to computational power limits, we simulated $M_B\sim 10$. The superscript $*$ denotes simulations with a copper shell as a boundary for the experimental low-temperature environment.}
\label{tab:comparison}
\end{center}
\end{table}

The axion-induced magnetic field $B_a$ can be measured by a SQUID-based magnetometer. The sensitivity of such a device is approximately $\Delta B\approx B_0*\sqrt{\Delta f/{\rm Hz}}$ Tesla, where $\Delta f$ is the bandwidth of the signal of interest. Due to Eq.(\ref{df}) and the fact that the low mass axions allow a larger $R$, the proposed scheme is more suitable for lighter axions.

$B_a$ permeates the inner part of the capacitor; thus, there are several ways to boost the signal. One can put a superconductor ring-coil inside the inner region. By using superconductors, the added ring-coil system is inductance dominated. The inductance of the ring is $L_r= R_rc_r$, where the inductance coefficient ($\mu_0=1$) $c_r\approx {\rm ln}(16R_r/d_r)-2$. $d_r$ is the diameter of the ring wire. The inductance of the coil is $L_c= R_cc_cN^2$, where $c_c$ is the coil's inductance coefficient and $N$ is its winding number. Assuming that $R_r\gg R_c$, $L_c$ and the mutual inductance can be neglected. Thus the total inductance $L\approx L_r$. The magnetic field seen by the magnetometer with current a $I$ is
\bea
B_1\approx {N\over 2R_c}I=-{N \over 2R_c L_r}\Phi_a
\eea
Because $\phi_a=\pi R_r^2B_a$, by adjusting the parameters, we can achieve a large enhancement. Please see Fig.(\ref{fig:pickup2},\ref{fig:pickup}) for an illustration.
The magnetic field inside the coil is then
\be
B_1\approx -{\pi R_rN \over 2R_c c_r}B_a=-F_rN{R_r\over R_c}B_a=-M_BB_a~,
\ee
where, $F_r\approx \pi/2c_r\sim {\cal O}(0.1)$ for ordinary cases. We use $M_B$ to denote the total boost factor, which can be on the order of $10^4$ for a modest setup.

\section{sensitivity forecast}

The noise of the experiment mainly comes from the thermal noise in the pickup and the magnetometer. With a superconducting pickup coil, the major source of white noise is the SQUID itself. Here, we will adopt the sensitivity in the frequency range dominated by the SQUID's white noise, where low-frequency noise is subdominant and the magnetic field is relatively frequency insensitive.
The capacity shells are enclosed in a low-temperature environment, and the silicon-type material, with a 1.1 eV energy gap between the valence band and the conduction band, suppresses thermal currents and unwanted screen effects compared with conductor shells. Typically, the temperature of the ring-coil system to be a superconductor is approximately several Kelvin, which should be sufficient for the proposed scheme. Certainly, read-out systems such as SQUID sensors could require a lower local working temperature.
\begin{figure}[t]
\includegraphics[scale=0.5]{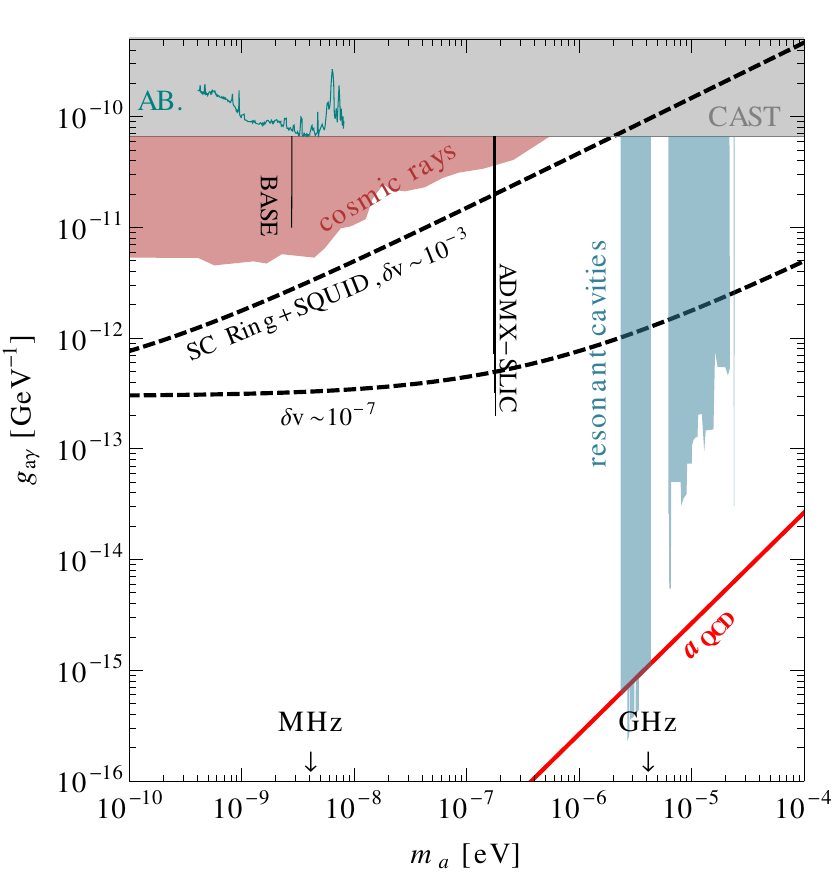}
\caption{The sensitivity of the axion-photon coupling $g_{a\gamma}$ of the proposed experiment with a preliminary setup $\bar E_0\sim 80$MVm$^{-1}$, $R\sim 1$m, $\delta v\sim 10^{-3}$, assuming 10 aT magnetic field sensitivity, and $M_B=10^4$. The limit for a more coherent DM distribution $\delta v\sim 10^{-7}$ is also shown. For comparison, we show the limits of CAST (gray), the collective resonant-cavity haloscope results (green) from ADMX, CAPP, HAYSTAC, etc., plus the recent lower-frequency experimental results from BASE~\cite{Devlin:2021fpq}, ADMX-SLIC~\cite{Crisosto:2019fcj} and ABRACADABRA (AB.)~\cite{Salemi:2021gck}. The red solid line $a_{\rm QCD}$ denotes the prediction from the KSVZ QCD-axion model.
}
\label{fig:constraints}
\end{figure}

In addition to thermal noise, other potential perturbations to the signal include, for instance, electric leakage between the capacitor plates and environmental kinetic fluctuations. Some vibrations can be calibrated in situ \cite{matichard20151, matichard20152}. The leakage current may induce a nontrivial background, but it should be independent of the direction of axion dark matter flow; therefore, one can distinguish leakage noise by adjusting the direction of the capacitor. The directional dependence of the signal is also helpful for reducing the vibration background by a factor of ${\cal O}(1)$ if several detectors with different orientations are operating simultaneously. Another major concern could be the breakdown electric-field strength of the capacitor. The electrical breakdown of silver or silver-nickel alloy in a 1.4$\time10^{-4}$ Pa vacuum is approximately 2-4$\times10^{9}$~V/m with negligible leakage current~\cite{zouache}. For semiconductors, Ref.~\cite{Lin2005} reported that the breakdown field strength of ultrathin atomic layers could reach 30 MV/cm. However, these experiments are in the microscopic region. In the near future, an electric field strength of approximately 0.1 GV/m \cite{DElia:2011fih} could be more reasonably implemented in the required macroscopic region.

Modern SQUID-based magnetometry has been developed for atto-Tesla detection and has reached magnetic field sensitivity~\cite{Degen:2017}
\bea
\Delta B \sim 10^{-16}{\rm ~Tesla}\cdot\sqrt{\Delta f/{\rm Hz}}+\Delta B_{\rm min}~~,
\label{eq:Bsensitivity}
\eea
where $\Delta f$ is the bandwidth of the measurement. Since our $\Delta f$ can be small towards lower $m_a$, we added $\Delta B_{\rm min}=10$ aT to denote an effective sensitivity cutoff in line with modern magnetic sensors. The local dark matter density $\rho_{\rm DM}\approx 0.4$ GeV~cm$^{-3}$, the dark matter velocity $v \approx 10^{-3}$, and
for $\Delta B$ larger than $\Delta B_{\rm min}$, we have the sensitivity on $g_{a\gamma}$ as
\bea
g_{a\gamma} &=&1.7\times 10^{-13}{\rm GeV}^{-1} \left({1{\rm m}\over {R}}\right)\left({1{\rm GV/m}\over {\bar E_0}}\right)\left({10^4\over M_B}\right)\nonumber\\
&\cdot& \sqrt{{m_a\over 10^{-5}{\rm eV}} {\delta v\over 10^{-7}}}~~,
\label{eq:squid_benchmark}
\eea
assuming a preliminary setup $\bar E_0\sim 80$ MVm$^{-1}$, $R\sim 1$m. The observation bandwidth assumes the axion energy uncertainty $\Delta f= 2m_a v\delta v$. The search limits are plotted in Fig.~\ref{fig:constraints} for the current $\Delta B\sim$10 atto-Tesla magnetic field sensitivity (dashed lines) for a conventional DM velocity dispersion $\delta v\sim 10^{-3}$ and a more coherent $\delta v\sim 10^{-7}$ situation for comparison. The Search for low-mass axions is advantageous due to the narrower frequency bandwidth, yet the limit flattens toward low frequencies as $\Delta B$ approaches the experimental sensitivity limit $B_{\rm min}$.

\medskip

\section{Discussion}
\label{sect:discussion}
In this paper, we propose the use of a cylindrical capacitor-created static electric field to assist the axion dark matter, inducing an oscillating magnetic field that can be further amplified by adding a superconductor ring-coil system. The deployment of a static electric field with a cylindrical manifold can result in a low noise oscillating magnetic field with an increased field strength. The proposed experiment is more sensitive to the lighter axions, which could be useful for wave-like dark matter searches.

Because the effective current depends on the electric field vector across the axion flow vector, the magnitude of the signal can be modulated by adjusting the angle between the axion velocity and the axis direction of the capacitor. A 24-hour modulation is expected with a fixed capacitor orientation due to the Earth's rotation. When the axion flow is nonparallel to the capacitor axis, a rotational nonsymmetric (around the axis) effective current pattern is generated, which induces a magnetic field with a component parallel to the pickup ring disk; thus, the dominant part of the signal is due to the parallel part of the axion flow. The superconducting ring-coil system is a pure induction pickup and thus could offer signal magnification over a wide frequency range.

In addition to SQUID-based magnetometers, recent developments in spin-based precision measurements, such as \cite{Wolf2015}
show high potential in very sensitive magnetic field detection. Such cryogenic detectors of suitable size can be utilized for signal measurement and may stimulate relative searches.

\medskip
{\bf Acknowledgement}
We thank Dongning Zheng, Zhihui Peng, Yirong Jin, Man Jiao, Qingjin Xu and Yi Zhou for useful discussions.
This work is supported by the National Natural Science Foundation of China (11875148, 12150010) and by the Institute of High Energy Physics, Chinese Academy of Sciences (Y95461A0U2).


\bigskip

\appendix

\section{EM simulations}
\label{app:sim}
Here, we show the simulation results of the electric and magnetic fields for the axion frequency at 1 MHz and at 10 MHz with a copper shell as the boundary of the low-temperature environment (Figs.7, 8 and 9). Fig.9, the 10 MHz case shows a strong field in the region immediately next to the inner cylinder of the capacitor. We simulated the scenario with different frequencies, and found that this is likely due to the skin effect which is more apparent at higher frequencies. In our simulations, the phenomenon started to be visible at frequencies above 5 MHz.

\begin{figure*}
\includegraphics[scale=0.6]{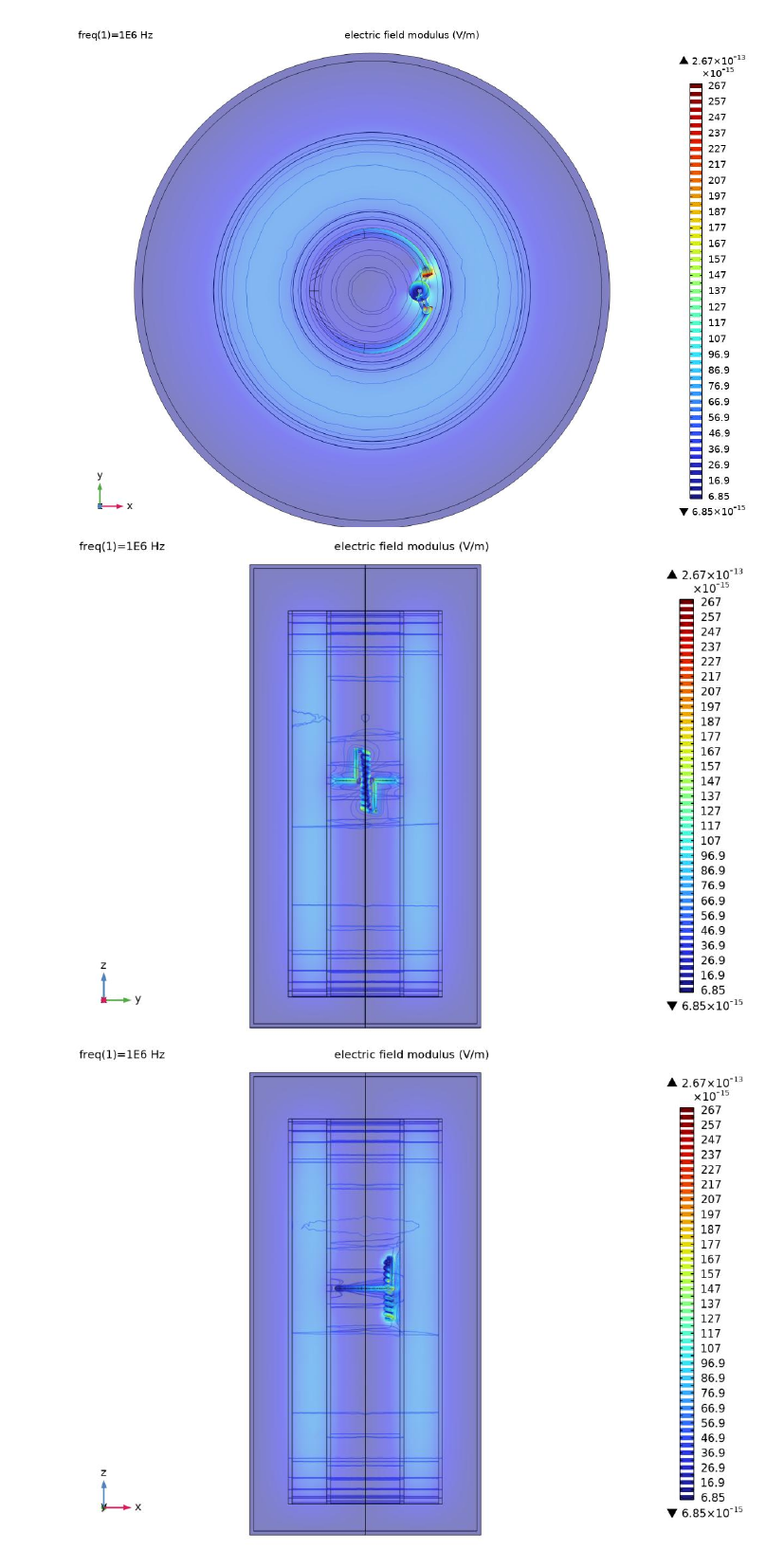}
\includegraphics[scale=0.6]{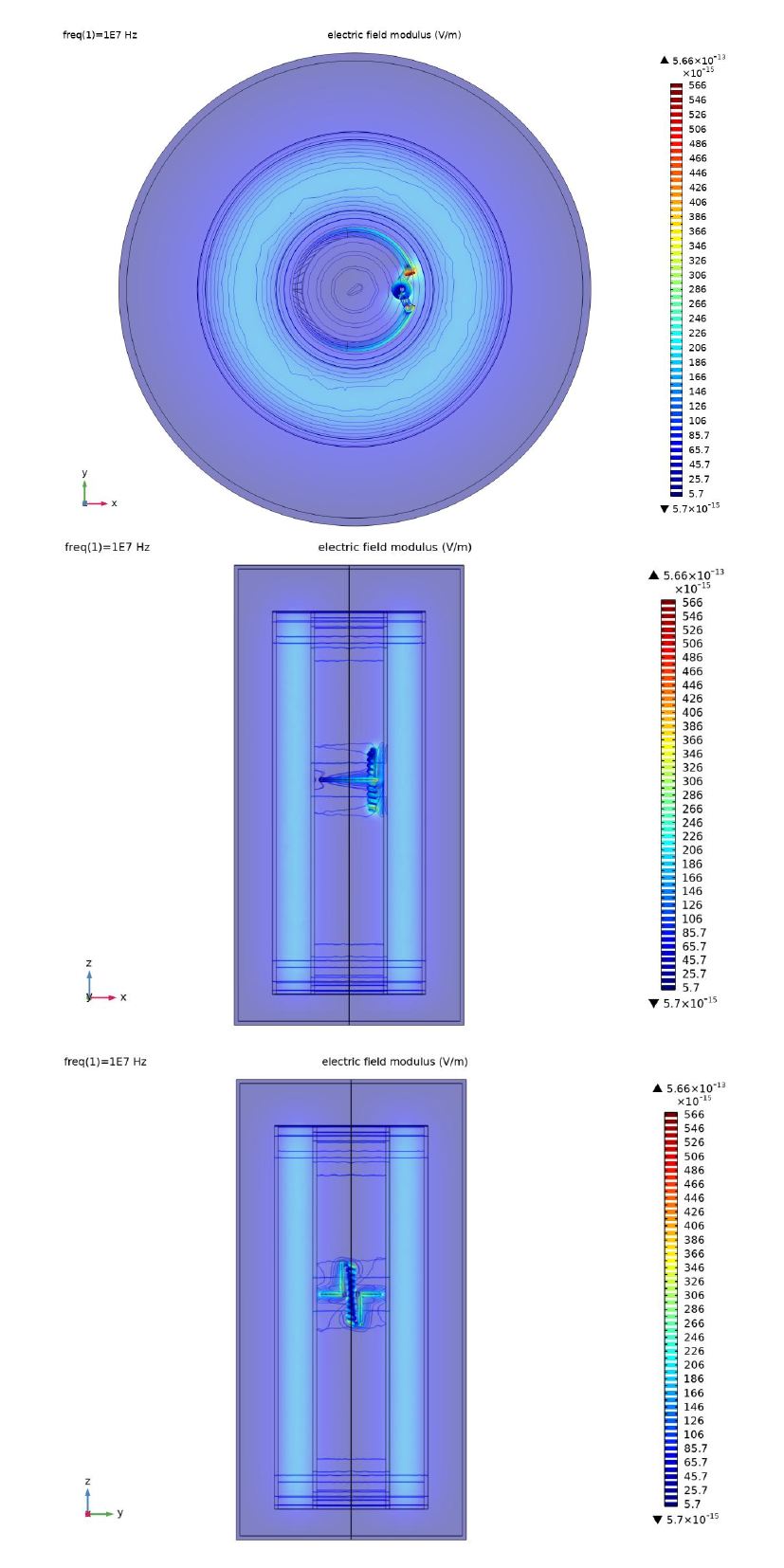}
\caption{Electric field distribution in the scheme with axion frequencies of 1 MHz (left) and 10 MHz (right).}
\label{fig:10MHz2}
\end{figure*}

\begin{figure*}
\includegraphics[scale=0.2]{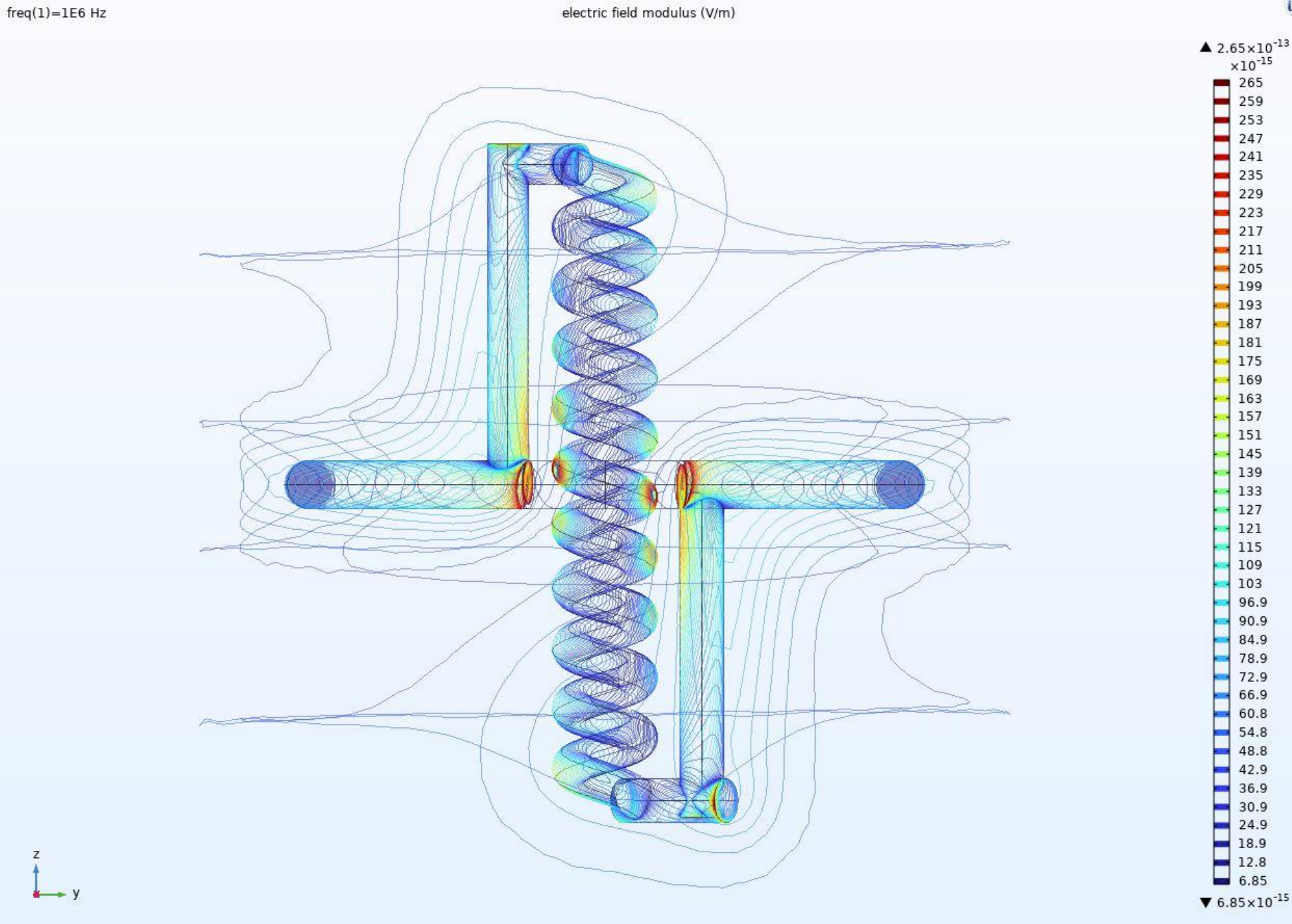}
\includegraphics[scale=0.2]{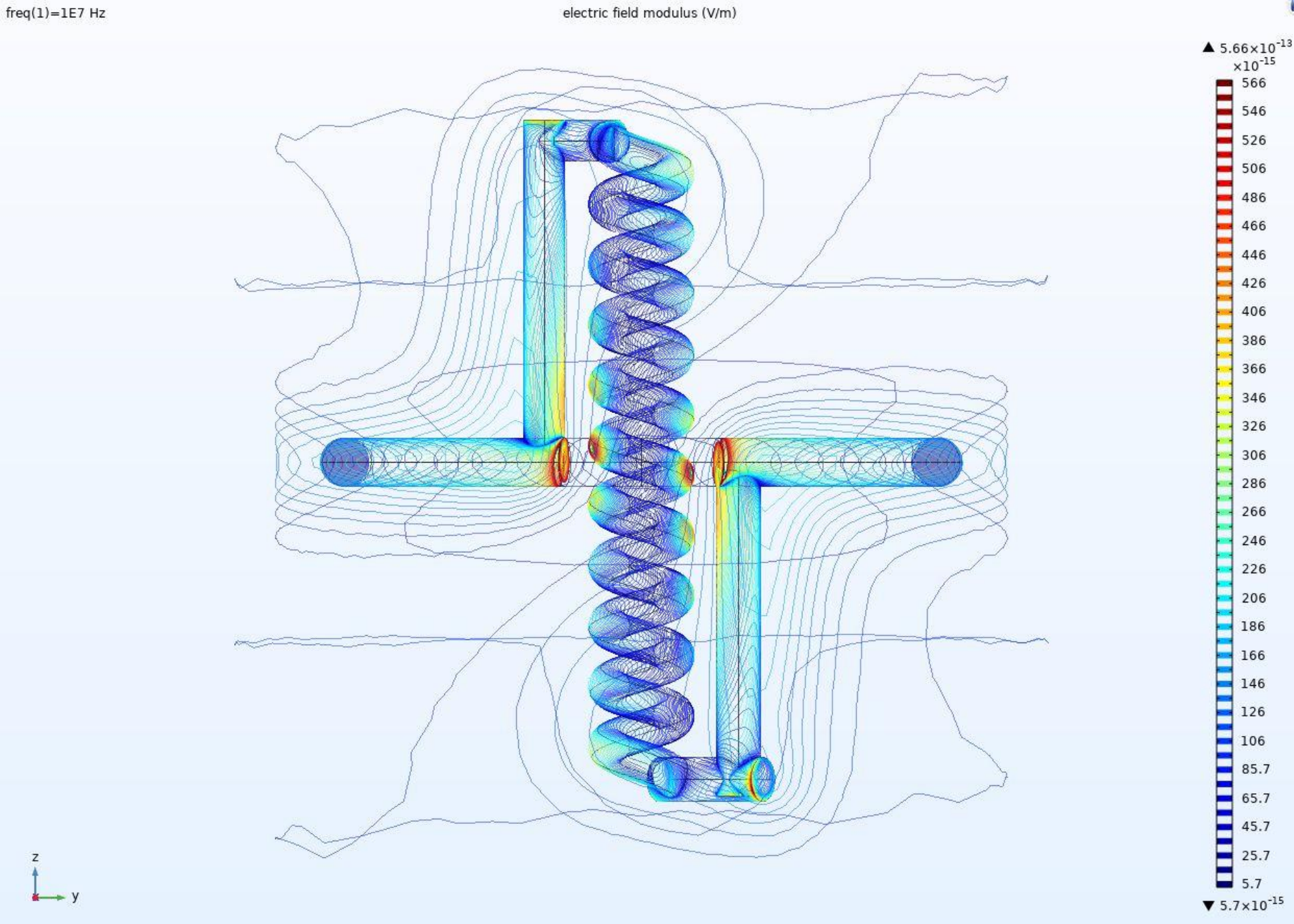}
\caption{Electric field distribution near the ring.}
\label{fig:10MHz1}
\end{figure*}

\begin{figure*}
\includegraphics[scale=0.2]{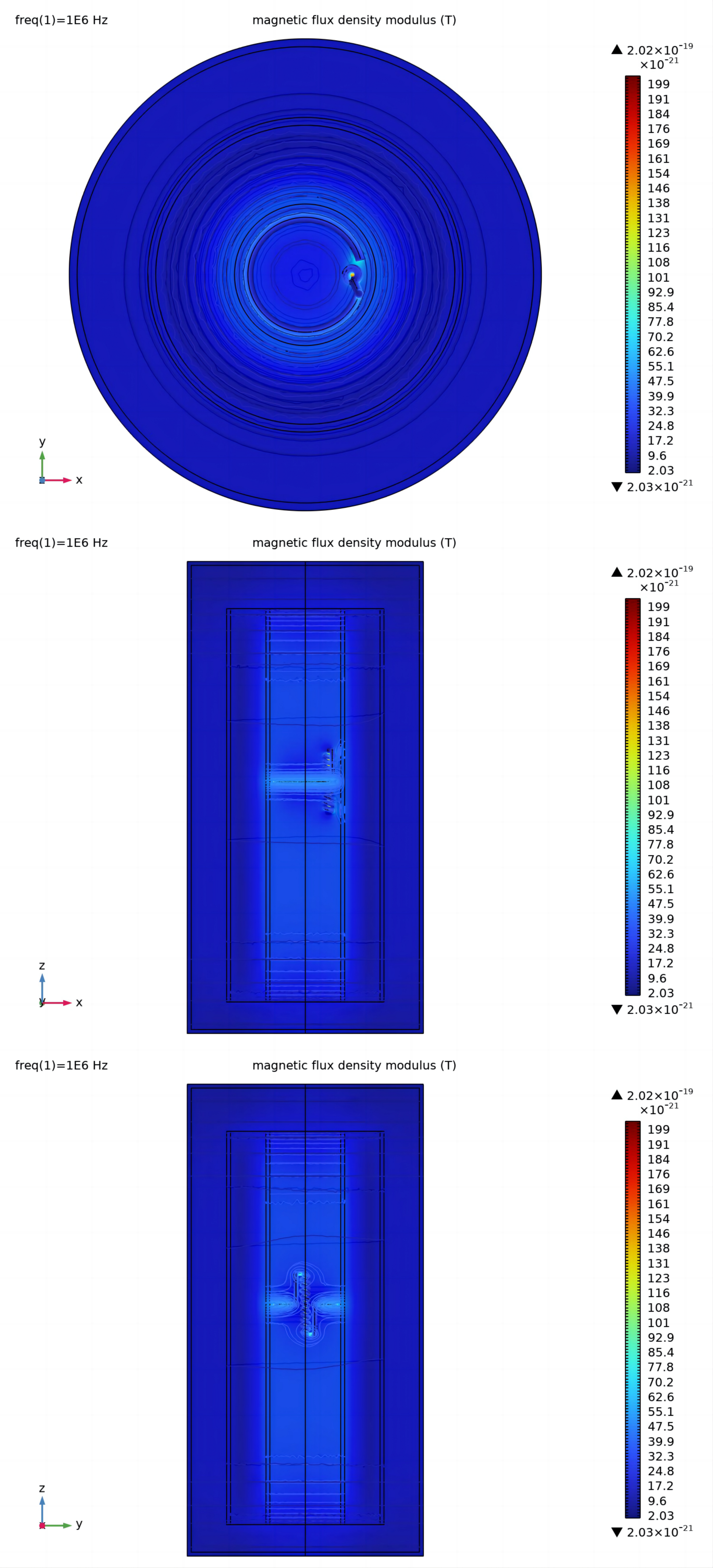}
\includegraphics[scale=0.2]{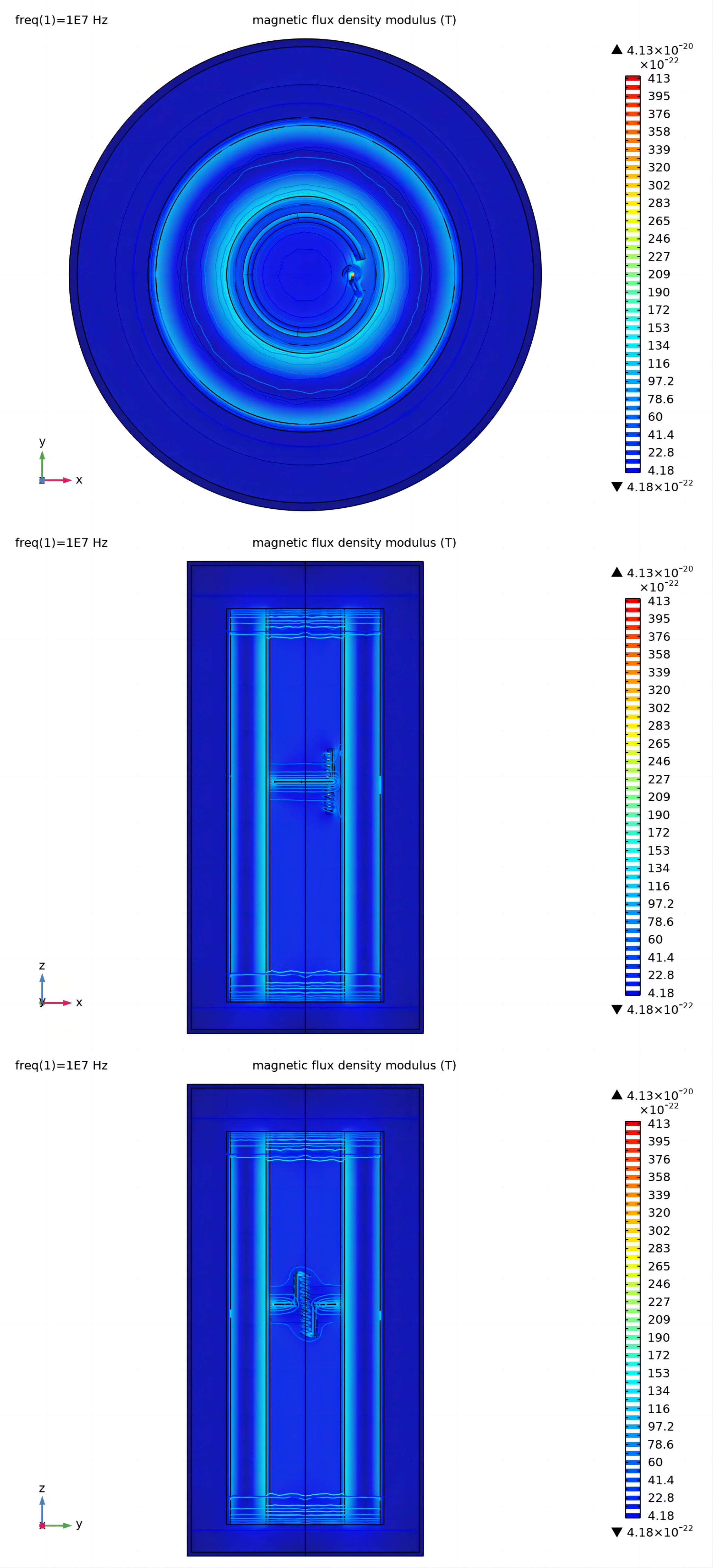}
\caption{Magnetic field distribution with axion frequencies of 1 MHz (left) and 10 MHz (right).}
\label{fig:10MHz1}
\end{figure*}

\end{document}